\documentclass[12pt,letterpaper]{article}
\input{epsf}
\usepackage[pdftex]{graphicx,color}
\usepackage{hyperref}
\input{psfig}
\usepackage{amsmath,amssymb}
\usepackage{fancyhdr}
\newcommand\ltap{\
  \raise.3ex\hbox{$<$\kern-.75em\lower1ex\hbox{$\sim$}}\ }
\newcommand\gtap{\
  \raise.3ex\hbox{$>$\kern-.75em\lower1ex\hbox{$\sim$}}\ }

\newcommand\simge{\mathrel{%
   \rlap{\raise 0.511ex \hbox{$>$}}{\lower 0.511ex \hbox{$\sim$}}}}
\newcommand\simle{\mathrel{
   \rlap{\raise 0.511ex \hbox{$<$}}{\lower 0.511ex \hbox{$\sim$}}}}

\newcommand{\slashchar}[1]%
        {\kern .25em\raise.18ex\hbox{$/$}\kern-.70em #1}
\def\lsim{\mathrel{\raise.3ex\hbox{$<$\kern-.75em\lower1ex\hbox{$\sim$}}}}
\def\gsim{\mathrel{\raise.3ex\hbox{$>$\kern-.75em\lower1ex\hbox{$\sim$}}}}
\newcommand{\bs}{\boldsymbol}

\newcommand\CL{{\cal L}}

\newcommand\CO{{\cal O}}

\newcommand\ul{\underline}

\newcommand\be{\begin{equation}}
\newcommand\ee{\end{equation}}
\newcommand\bea{\begin{eqnarray}}
\newcommand\eea{\end{eqnarray}}
\newcommand\ba{\begin{array}}
\newcommand\ea{\end{array}}

\newcommand{\thalf}{\textstyle{\frac{1}{2}}}

\newcommand\gev{{\rm GeV}}
\newcommand\tev{{\rm TeV}}

\newcommand\ellm{\ell^-}

\newcommand\ellp{\ell^+}

\begin{document}

\title{ \vspace{-50mm}
  {\Large{\bf The Composite Higgs Signal\\ at the Next Big 
      Collider}\footnote{This paper is dedicated to my friend and collaborator,
    Eric Pilon.}}\\
  \medskip } \author{Kenneth Lane\thanks{lane@bu.edu}\\
  {\large Department of Physics}\\{\large Boston University}\\
  {\large 590 Commonwealth Avenue}\\ {\large Boston, Massachusetts 02215}\\
} \maketitle

\vspace{-1.0cm}

\begin{abstract}

  The Gildener-Weinberg (GW) mechanism produces a Higgs boson $H$ that is
  a dilaton. That is, $H$ is both naturally light and naturally aligned. It
  also predicts additional singly-charged and neutral Higgs bosons all of
  whose masses are $\simle 500\,\gev$ and, therefore, within reach of the LHC
  {\em now}. I argue that the GW Higgs is composite --- a bound state of
  fermions whose strong interactions are at some high, unknown scale
  $\Lambda_H \simge 1\,\tev$. The lone harbingers of $H$ compositeness, ones
  that may be accessible at the next multi-TeV collider, are isovector vector
  $\rho_H$ and axial vector $a_H$ bound states whose masses are
  $\CO(\Lambda_H)$. They decay into the only fermion-antifermion composites
  lighter than they are, the Higgs boson and longitudinally-polarized weak
  bosons: $\rho_H^{\pm,0} \to W^\pm_{\CL} Z_{\CL}$, $W^+_{\CL} W^-_{\CL}$ and
  $a_H^{\pm,0} \to W^\pm_{\CL} H$, $Z_{\CL} H$. Observing these resonant,
  highly-boosted weak-scale bosons would establish their composite nature.

\end{abstract}


\newpage

\section*{1. Why I think the Higgs is composite}

No one believes that the 125-GeV Higgs boson $H$~discovered at CERN in
2012~\cite{Aad:2012tfa,Chatrchyan:2012ufa} is anywhere near all there is to
the Higgs sector. As a theoretical construct, $H$ has so many shortcomings
--- which hardly need repeating here\footnote{But to name the most serious, see
Refs.~\cite{Wilson:1970ag},\cite{'tHooft:1979bh}.} --- that they overshadow the
essential roles it plays in the Standard Model of breaking electroweak
symmetry and giving mass to the weak gauge bosons and (most) fermions. Thus,
the history of elementary particle physics since 1972 has been dominated by
the search for and proposal of solutions to these deficits.

\begin{figure}[ht!]
 \begin{center}
\includegraphics[width=2.65in, height=2.65in]{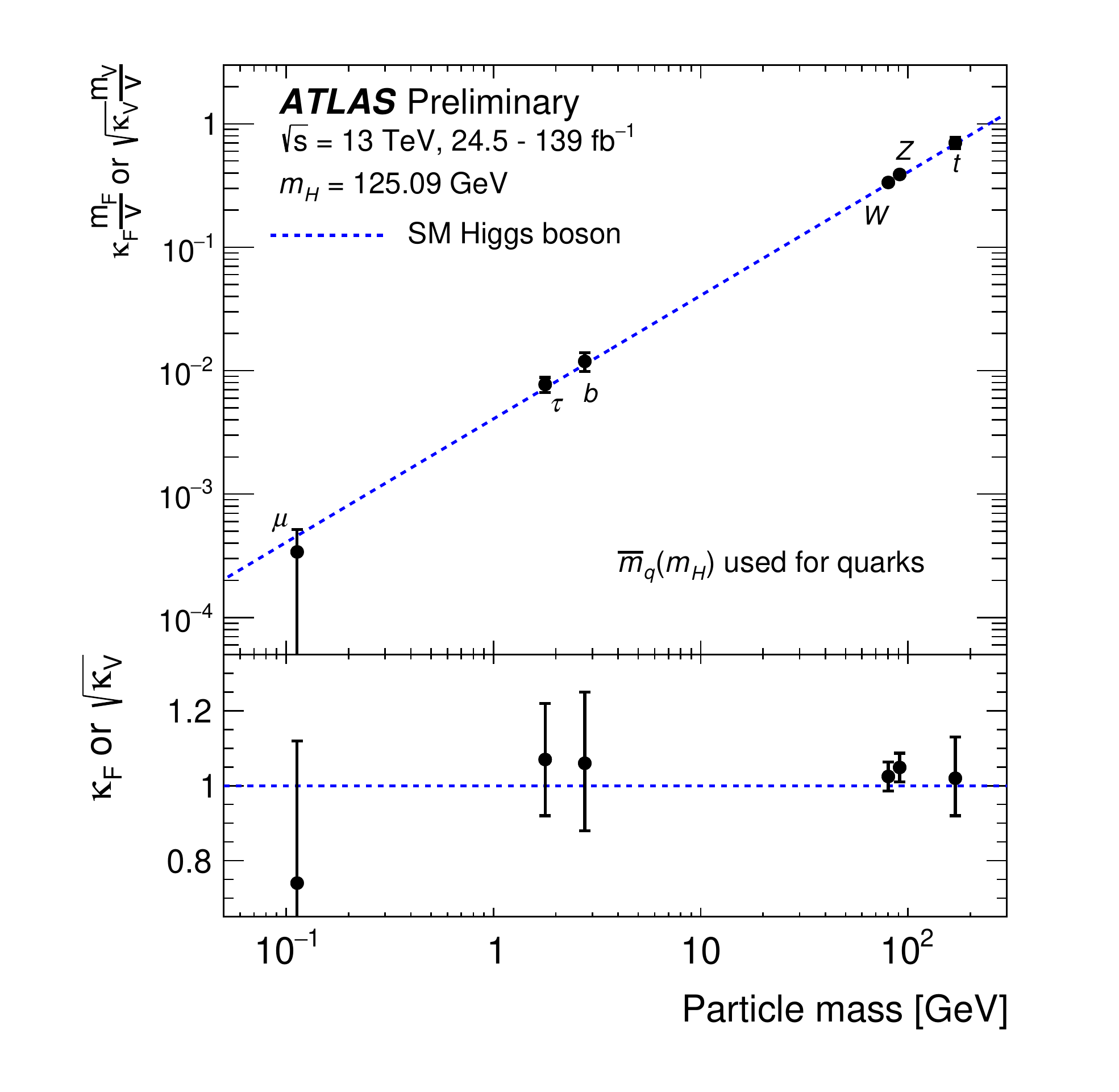}
\includegraphics[width=2.65in, height=2.65in]{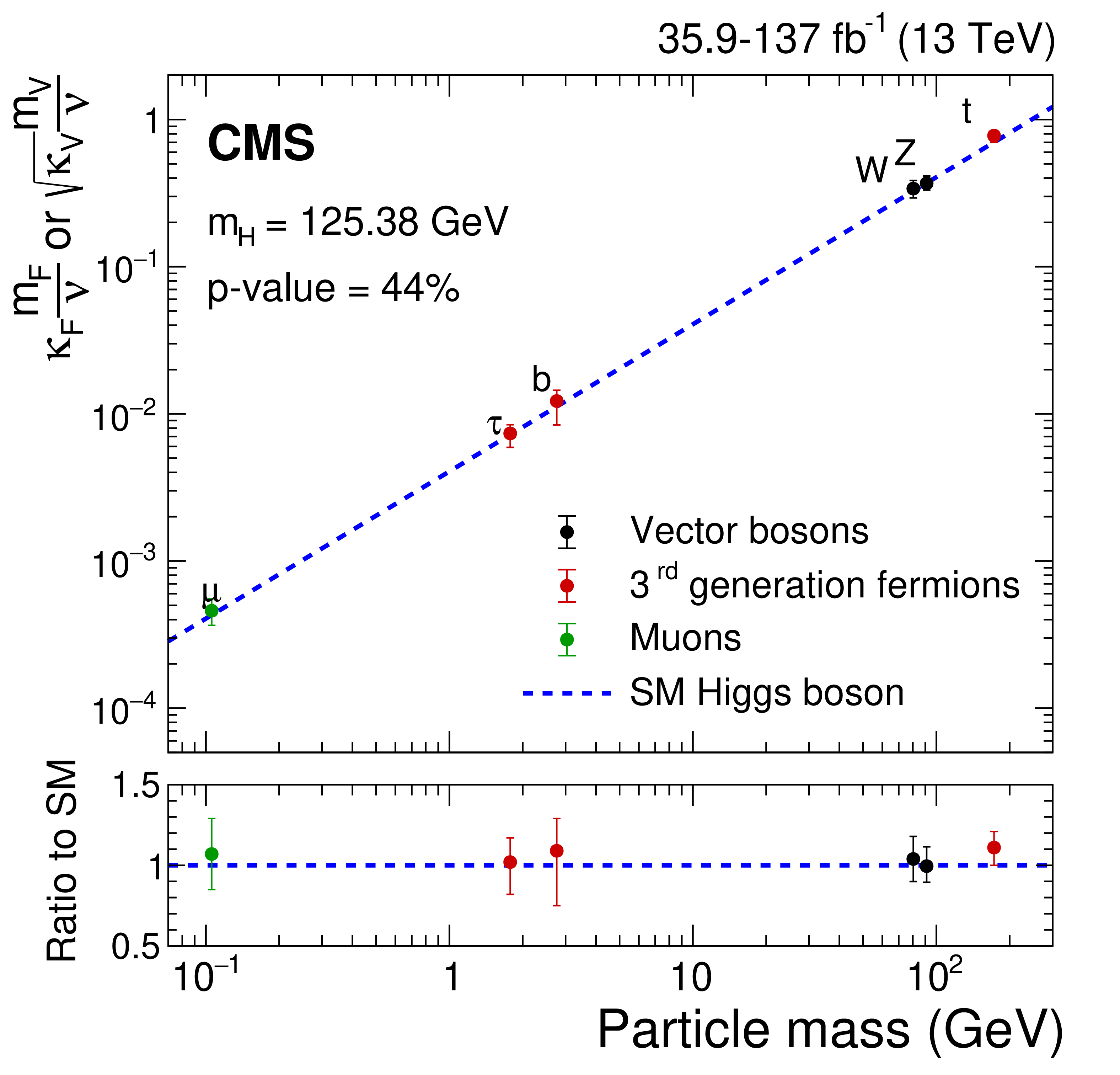}
\caption{The mass-dependent couplings of quarks, leptons and the $W$ and $Z$
  as measured by
  ATLAS~({\url{https://atlas.web.cern.ch/Atlas/GROUPS/PHYSICS/CombinedSummaryPlots/HIGGS/}})
  and CMS
  ({\url{https://cms-results.web.cern.ch/cms-results/public-results/publications/HIG/index.html}}).}
  \label{fig:SMHiggs}
 \end{center}
 \end{figure}

The solutions that have been proposed invariably require additional Higgs
bosons. The more popular of these include supersymmetry, little Higgs models,
extended weak gauge symmetries, dark sectors and, prosaically, multi-Higgs
doublet models which, often, are more or less well-motivated by overarching
theoretical constructs such as those just mentioned. After all these years,
however, and especially after all the heroic searches for extensions of the
Standard Model's $SU(2)\otimes U(1)$ gauge symmetry~\cite{Glashow:1961tr} and
its single complex Higgs doublet, there is {\em no evidence} that the Higgs
boson is anything other than that proposed so long
ago~\cite{Weinberg:1967tq}. Not only are there no extra Higgs bosons, there
are no TeV-scale partners of the top quark and the weak $W$ and $Z$ bosons,
there are no Higgsinos, squarks, sleptons, gaugeinos, no experimental support
for dark portals such as long-lived particles, no sign of vector-quarks or
vector-leptons, nothing new at all since 2012. And, to belabor the point, the
Higgs $H$ appears in every respect to be that expected in the Standard Model:
all measurements so far of its interactions with weak bosons and massive
fermions are within one standard deviation of the Standard Model's
predictions; see Fig.~\ref{fig:SMHiggs}.

Yet, we still believe there is more to the Higgs than the Standard Model
(SM). A major difficulty of this belief is that, if there are other Higgs
bosons, why should exactly one mass eigenstate scalar have SM couplings? The
usual answer is ``Higgs
alignment''~\cite{Boudjema:2001ii,Gunion:2002zf,Carena:2013ooa,
  Haber:2018ltt}. However, alignment often assumes a sizable hierarchy of
Higgs masses so that the lightest Higgs decouples and has SM couplings. With
a few exceptions that rely on elaborate global symmetries or
supersymmetry (see, e.g., Refs.~\cite{Ivanov:2007de,Dev:2014yca,
  Benakli:2018vqz,Darvishi:2019ltl}), such decoupling suffers from large
radiative corrections.

 Another possibility is that the Higgs boson is composite. This was always
 the case in technicolor, but there was no obvious reason why the Higgs would
 be much lighter than the technicolor scale of $\CO(1\,\tev)$.

 The way out of this is that $H$ is a dilaton, a pseudo-Goldstone boson of
 spontaneously broken scale invariance that is also explicitly broken at some
 scale~$f$. See Ref.~\cite{Lane:2018ycs} for some earlier references on this
 subject. A major advantage of the dilaton scheme is that its couplings to
 weak bosons and fermions have the same form as the SM Higgs'. However, those
 couplings are proportional to $f$, and $f \neq v = 246\,\gev$ in
 general. There is one exception to this: $f = v$ if {\em only} operators
 that are charged under the electroweak gauge group have
 conformal-symmetry-breaking vacuum expectation values, i.e., if the agent
 responsible for electroweak symmetry breaking, the Higgs boson, is also the
 one responsible for explicit scale symmetry
 breaking~\cite{Bellazzini:2012vz}.

 Now, this is an intriguing possibility and one that was realized long
 ago~\cite{Gildener:1976ih} yet not generally recognized as such. As I'll
 argue next, I believe this possibility makes sense only if $H$ is composite.

\section*{2. Why I think ${\bs H}$ is the Gildener-Weinberg
  dilaton} 

If $H$ is the massless dilaton of spontaneously-broken scale symmetry, its
low-energy Lagrangian must be classically (i.e., at tree level)
scale-invariant. How this happens is a mystery. As far as we understand, the
responsibility lies with scale-invariant interactions of massless fermions at
some higher energy scale $\Lambda_H$. Presumably, these are strong gauge
interactions (S.I.) that generate $H$ as a bound state of the
fermions~\cite{'tHooft:1979bh}. One thing we know about these interactions is
that
\be\label{eq:lamH}
  \Lambda_H \simge 1\,\tev.
\ee

The S.I.~fermions must transform under electroweak $SU(2)\otimes U(1)$ so
that $H$ does and, therefore, have weak isospin ${\thalf}$ and $0$. Assuming
that their chiral symmetry contains the electroweak symmetry, these fermions
must also produce the three Goldstone bosons, $W^\pm_{\CL}$ and $Z_{\CL}$,
that become the longitudinal ($\CL$) components of the electroweak gauge
bosons. The four massless bound states $(H,W^\pm_{\CL},Z_{\CL})$ then form a
complex $(2_L,2_R)$ doublet $\Sigma$ under $SU(2)_L\otimes SU(2)_R$ which is
contained in the S.I.~fermions' chiral symmetry. The low-energy theory also
contains quarks and leptons and, possibly, other scalars. They too must be
massless at tree-level to maintain the scale-invariance. This is natural for
the known fermions~$\psi$ since they transform as left-handed electroweak
doublets and right-handed singlets. Then, $\Sigma$~couples $\psi_L$ to
$\psi_R$ to break their chiral symmetry. If there are additional scalars,
their self-interactions must be purely quartic, as are their Yukawa and gauge
interactions. This must be enforced by the scale-invariant S.I. at
$\Lambda_H$.

Finally, the scale invariance must be explicitly broken so that all these
massless particles including $H$ (but not the photon) acquire mass. This can
happen as a consequence of the renormalization of the low-energy theory with
the appearance of a massive renormalization scale. This is the mechanism of
S.~Coleman and E.~Weinberg for generating masses in a (classically)
scale-invariant theory~\cite{Coleman:1973jx}.

\subsection*{2.a The Gildener-Weinberg 2HDM}

The low-energy effective Lagrangian for this picture was written down
46~years ago by E.~Gildener and S.~Weinberg
(GW)~\cite{Gildener:1976ih}. Their aim was to use it to produce a very light
Higgs boson. In doing this, GW assumed that all quarks and leptons are light
compared to the weak scale (which, of course, they were then) and that all
the quartic scalar self-interactions were of order $e^2$. (They couldn't be
smaller than that because of electroweak radiative corrections.) We now know
that the top quark is very heavy and, so, it turns out there is no need for
the second assumption on the scalar self-couplings. See Eq.~(\ref{eq:MHsq})
for the need for heavier scalars in the presence of the top quark.

GW did not adopt a specific model of their scheme. However, using the
Coleman-Weinberg expansion for the low-energy effective potential, they
derived a very important formula for the Higgs mass:
\be\label{eq:MHsq} M_H^2 = \frac{1}{8\pi^2 v^2}\left(3\sum_{V}M_V^4 +
  \sum_{S}M_S^4 - 4\sum_{F} M_F^4)\right), \ee
where the sums are over the degrees of freedom (polarizations, colors, etc.)
of massive gauge bosons~$V$, scalars~$S$ and fermions~$F$.

As already alluded to above, another very important consequence of
Ref.~\cite{Gildener:1976ih} is that the $H$ couplings to fermions and weak
gauge bosons have the same form as in the SM. When the conformal symmetry is
explicitly broken in the one-loop potential, all those couplings are
proportional to the Higgs vacuum expectation value, $v = 246\,\gev$. That is,
this Higgs is {\em aligned!} The one-loop corrections to perfect alignment
are very small and would be absent altogether were it not for the top
quark~\cite{Eichten:2021qbm}. Thus, the alignment is natural and the
departures from perfect alignment naturally small.

The simplest model employing the GW scheme was proposed by Lee and Pilaftsis
in 2012~\cite{Lee:2012jn}. The model assumes the standard electroweak gauge
symmetry with the known quarks and leptons. It also has two Higgs doublets so
that, in addition to $\Sigma = (H,W^\pm_{\CL},Z_{\CL})$, the second doublet
is $\Sigma' = (H',H^\pm,A)$ where $H'$ is $C\!P$-even and $A$ is
$C\!P$-odd.\footnote{The quartic scalar potential is automatically
  $C\!P$-conserving~\cite{Lee:2012jn,Lane:2018ycs}.} Because of the
dominant role the Gildener-Weinberg mechanism plays in this model, I refer
to it as the GW-2HDM.

\subsection*{2.b What are the signals of the GW-2HDM?}

With $M_H = 125\,\gev$, Eq.~(\ref{eq:MHsq}) implies a sum rule for the masses
of the new Higgs bosons:
\be\label{eq:MHsum}
\left(M_{H'}^4 + M_A^4 + 2M_{H^\pm}^4\right)^{1/4} = 540\,\gev.
\ee
This sum rule holds in the one-loop approximation of {\em {\ul{any}}} GW
model of electroweak symmetry breaking in which the only weak bosons are $W$
and $Z$ and the only heavy fermion is the top quark. Thus, the larger the
Higgs sector, the lighter will be the masses of at least some of the BSM
Higgs bosons expected in a GW model. In short, these models predict the new
Higgs bosons at surprisingly low masses.

These light BSM Higgs bosons are by far the surest way to to test the GW-2HDM
at the LHC in this decade and, perhaps, for longer than that. The current
experimental situation is summarized in Refs.~\cite{Lane:2019dbc,
  Eichten:2021qbm}. To avoid conflict with precision measurements of the
$T$-parameter, $M_{H^\pm} = M_A$ is assumed (see Ref.~\cite{Lee:2012jn} and
references therein). Then, $M_{H'}$ can be taken from the sum
rule~(\ref{eq:MHsum}). The 2HDM parameter $\tan\beta \simle 0.50$ for
$180\,\gev < M_{H^\pm} < 550\,\gev$.\footnote{The version of the GW-2HDM
  discussed in Refs.~\cite{Lane:2018ycs,Lane:2019dbc, Eichten:2021qbm} has
  the structure of the usual type-I 2HDM~\cite{Branco:2011iw}, but with the
  Higgs doublets $\Phi_1$ and $\Phi_2$ interchanged. The effect of this is
  that experimental lower limits on $\tan\beta = v_2/v_1$ in other type-I
  models are lower limits on $\cot\beta$ in this model.} This limit comes
from a CMS search with $8\,\tev$ data for
$H^+ \to t\bar b$~\cite{Khachatryan:2015qxa}. Subsequent searches by
ATLAS~\cite{Aaboud:2018cwk} and CMS~\cite{Sirunyan:2019arl} at $13\,\tev$
have not improved on this limit because the $t\bar t$ background is large and
grows faster with energy than the signal.\footnote{Above
  $M_{H^\pm} = M_A \simeq 400\gev$, the sum rule implies such a light
  $M_{H'}$ that it decays to $b\bar b$ or two gluons, a signal that is
  overwhelmed by the QCD background.}

Similar low-energy difficulties afflict other searches. The decay $A$ or
$H' \to \bar tt$ at and not far above the $\bar tt$ threshold at $350\,\gev$
are subject to theoretical uncertainties in the QCD production rate
there~\cite\cite{Sirunyan:2019wph}. For lower mass $A$ or $H'$, their decays
to $\bar bb$ are swamped by the QCD backgrounds.

There have been many other LHC searches for BSM Higgs bosons, almost
exclusively at higher masses. Two examples are $gg$ or weak-boson fusion of
$H'$ and $A$ followed by their decay to $ZH$ or to $WW$ and $ZZ$. These and
many other searches have been fruitless. That is expected for the GW-2HDM
(and similar models). Many if not most of these searches have been based on
processes that are forbidden for an aligned
Higgs~$H$.\footnote{\url{https://twiki.cern.ch/twiki/bin/view/AtlasPublic}}$^,$
\footnote{\url{htps://cms-results.web.cern.ch/cms-results/public-results/publications/HIG/SUS.html}}

\section*{3. What role can the next big collider play?}

Although the low-mass signals of the GW-2HDM are well within reach of the LHC
with its 13-14~TeV collision energies and high luminosities, they are not
accessible to the ATLAS and CMS detectors because of their difficulty
overcoming the QCD backgrounds at such masses. It is to be hoped that the
detector and analysis improvements being made for Run~3 will remedy this.

However, there is one signal of these models that must exist somewhere above
$1\,\tev$ and which is probably GW-model independent. That is the existence
of heavy spin-one bound states of the S.I.~fermions. They have an isospin
$I = 1$ (and $0$) inherited from the weak isospin of the fermions. Their
masses are $\CO(\Lambda_H)$ and, so, unknown. But we look where we can, and
the planners for the next big collider being discussed in Europe, the US and
China would do well to make searching for these resonances a priority. They
and the way they decay will be direct evidence that $H$ and
$W^\pm_{\CL},\,Z_{\CL}$ are composites of the S.I.~fermions.

The S.I. have a parity-invariance, much like the parity of QCD. Because of
the parity inherent in the $(2_L,2_R)$ symmetry of $\Sigma$, the isovector
bosons will be (ordinary) vectors and axial vectors analogous to the $\rho$
and $a_1$ of hadron physics. Unlike QCD, they are expected to be nearly
degenerate. I will call them $\rho_H$ and $a_H$ to emphasize their connection
to the Higgs $H$. For more theoretical background and details of the S.I. and
their symmetry structure, see Refs.~\cite{Appelquist:2015vdl,Lane:2015fza,
  Lane:2009ct}.

The $\rho_H$ and $a_H$ are produced mainly by the Drell-Yan process:
\bea\label{eq:DY}
\bar q' q &\to& W^\pm, Z, \gamma \to
\rho_H^{\pm,0},\,\,a_H^{\pm,0}\,\,\,\,\hspace{0.07cm}{\text{in a hadron collider;}}\\
\ellp\ellm &\to& Z,\gamma\to \rho_H^0,\,\,a_H^0 \,\,\hspace{1.50cm}{\text{in
    a lepton collider.}}
\eea
In a $pp$ collider, there is also weak-boson fusion (VBF) of $\rho_H$ and
$a_H$. At the LHC, VBF accounts for only 20\% of $\rho_H$ production and very
little for $a_H$~\cite{Lane:2015fza}. This fraction needs to be determined at
much higher energies. For an $\ellp\ellm$ collider, the greatest reach,
perhaps competitive with a $100\,\tev$ $pp$ collider, might be achieved by a
muon collider.\footnote{I thank Tulika Bose for making this point.} This
would be an interesting study for the Snowmass Muon Collider Forum.

The principal decays of $\rho_H$ and $a_H$ are to the only S.I. fermion bound
states lighter than themselves, namely, the dilaton $H$ and the longitudinal
weak bosons $W^\pm_{\CL},\,Z_{\CL}$ --- the ``pions'' of S.I.~physics. These
decays obey the parity of the S.I.~interactions.\footnote{This is {\em not}
  the often benchmarked HVT model!}
\bea
\label{eq:rHdecay}
\rho_H^\pm &\to& W^\pm_{\CL}Z_{\CL},\qquad \rho_H^0 \to
W^+_{\CL} W^-_{\CL} \qquad({\rm but\,\,not\,\,to}\,\,Z_{\CL}Z_{\CL});\\
\label{eq:aHdecay}
a_H^\pm &\to& W^\pm_{\CL} H, \qquad\,\, a_H^0 \to Z_{\CL} H.
\eea
For $M^2_{\rho_H} \cong M^2_{a_H} \gg M^2_{W,Z,H}$ the final-state bosons are
highly-boosted and the longitudinal polarization vectors
$\epsilon_{\CL} \cong M_{\rho_H}/2M_{W,Z}$. This makes the otherwise
weak-decay rates of 
Eqs.~(6,7)strong. They are~\cite{Lane:2015fza}
\bea\label{eq:rates}
\Gamma(\rho_H^0 \to W^+ W^-) &\cong& \Gamma(\rho_H^\pm \to W^\pm Z) \cong
 \frac{g_{\rho_H}^2 M_{\rho_H}}{48\pi},\\
 \Gamma(a_H^0 \to ZH) &\cong& \Gamma(a_H^\pm \to W^\pm H) \cong
 \frac{g_{a_H}^2 M_{a_H}}{48\pi},
\eea
where the S.I.~couplings $g_{\rho_H} \cong g_{a_H} = \CO(1)$
(presumably). Then, $\Gamma(\rho_H,a_H) = \CO(M_{\rho_H}/100)$.  Finally,
since the $\bar qq'$ or $\ellp\ellm$ annihilation to $\rho_H$ and $a_H$
occurs with one unit of angular momentum along the beam axis, the decay
bosons will be emitted with a $\sin^2\theta$ angular distribution in the
$\rho_H/a_H$ rest frame.

\section*{Acknowledgments}

It is a pleasure to acknowledge my collaborators Estia Eichten, Eric Pilon,
Lukas Pritchett and Will Shepherd. I also thank Tulika Bose and Kevin Black
for comments and suggestions. 


\bibliography{Composite-Higgs_EF09}
\bibliographystyle{utcaps}
\end{document}